\documentclass{ws-procs975x65}

\usepackage{mathrsfs}

\begin{document}

\title{DISTORTION OF NEUTRON STARS \\ WITH A TOROIDAL MAGNETIC FIELD}

\author{J.~FRIEBEN and L.~REZZOLLA}

\address{Max-Planck-Institut f\"ur Gravitationsphysik
  (Albert-Einstein-Institut) \\
  D-14476 Golm, Germany}

\begin{abstract}
Models of rotating relativistic stars with a toroidal magnetic field
have been computed for a sample of eight equations of state of cold
dense matter. Non-rotating models admit important levels of
magnetization and quadrupole distortion accompanied by a seemingly
unlimited growth in size. Rotating models reach the mass-shedding
limit at smaller angular velocities than in the non-magnetized case
according to the larger circumferential equatorial radius induced
by the magnetic field.
Moreover, they can be classified as prolate--prolate, oblate--prolate,
or oblate--oblate with respect to surface deformation and quadrupole
distortion. Simple expressions for surface and quadrupole deformation
are provided that are valid up to magnetar field strengths and rapid
rotation.
\end{abstract}


\bodymatter

\section{Introduction}\label{sec:intro}

Neutron stars with a strong toroidal magnetic field have attracted
increasing interest as the magnetically induced distortion of
their matter distribution may lead to the quasi-periodic emission
of gravitational waves~\cite{Cutler2002,Bonazzola1996}, for
example, in the case of low-mass X-ray binaries (LMXBs).
Moreover, strong magnetic fields are believed to power the
electromagnetic activity of magnetars, which subsume both anomalous
X-ray pulsars (AXPs) and soft-gamma repeaters
(SGRs)~\cite{Duncan1992,Thompson1996}. Models of relativistic stars
with a toroidal magnetic field can be obtained within the standard
formalism for stationary and axisymmetric relativistic
stars~\cite{Bonazzola1993}, since the electromagnetic
stress--energy tensor then satisfies the same compatibility
condition~\cite{Oron2002} as the stress--energy tensor of an
unmagnetized perfect fluid
in purely rotational motion. Based on this finding, numerical models
of relativistic stars with a toroidal magnetic field have
emerged~\cite{Kiuchi2008,Frieben2012} whereas the poloidal case
was already studied a long time ago~\cite{Bocquet1995}.

\section{Method and results}
The neutron star matter is modeled as a perfectly-conducting perfect
fluid at zero temperature, described by a one-parameter equation of
state (EOS). For stationary and axisymmetric models in rigid rotation
as considered hereafter, the
general-relativistic line element in spherical coordinates
$(t,r,\theta,\phi)$ can be chosen as
\begin{equation} \label{e:ds2}
   \mathrm{d}s^2 = - N^{2} \mathrm{d}t^{2} + \mathit{\Phi}^{2} r^{2}
   \sin^{2} \theta^{2} ( \mathrm{d}\phi - N^{\phi} \mathrm{d}t )^{2}
   + \mathit{\Psi}^{2} ( \mathrm{d}r^{2}
   + r^{2} \mathrm{d}\theta^{2} )
\end{equation}
with gravitational potentials $N$, $N^{\phi}$, $\mathit{\Psi}$, and
$\mathit{\Phi}$ that are functions of $(r,\theta)$ alone.
The toroidal magnetic field must then ensure that the Lorentz force is
the gradient of a scalar potential, which is the case for
$B = \lambda_{0} \, (e + p) \, \mathit{\Phi}
N r \sin\theta$, where $e$ is the proper energy density of the
fluid, $p$ is the fluid pressure, and $\lambda_{0}$ is the
magnetization parameter. The field and matter equations are derived
from the perfect-fluid case~\cite{Bonazzola1993} by taking into
account additional magnetic source terms, expressed in terms of $B$,
and the magnetic potential $\tilde{M} = \lambda_{0}^{2} / (4\pi) \,
(e + p) \mathit{\Phi}^{2} N^{2} r^{2} \sin^{2} \theta$, supplemented
by the above relation for $B$ and the EOS.

The numerical models have been computed by means of a multidomain and
surface-adaptive pseudo-spectral code for stationary and axisymmetric
relativistic stars from the
\textsc{lorene}\footnote{\url{http://www.lorene.obspm.fr}} package,
extended to the case of the toroidal magnetic field specified
above, and employing its standard sample of nuclear matter EOSs.

\def\figsubcap#1{\par\noindent\centering\footnotesize(#1)}
\begin{figure}[t]
\begin{center}
  \hspace*{-0.2in}
  \parbox{2.0in}{\epsfig{figure=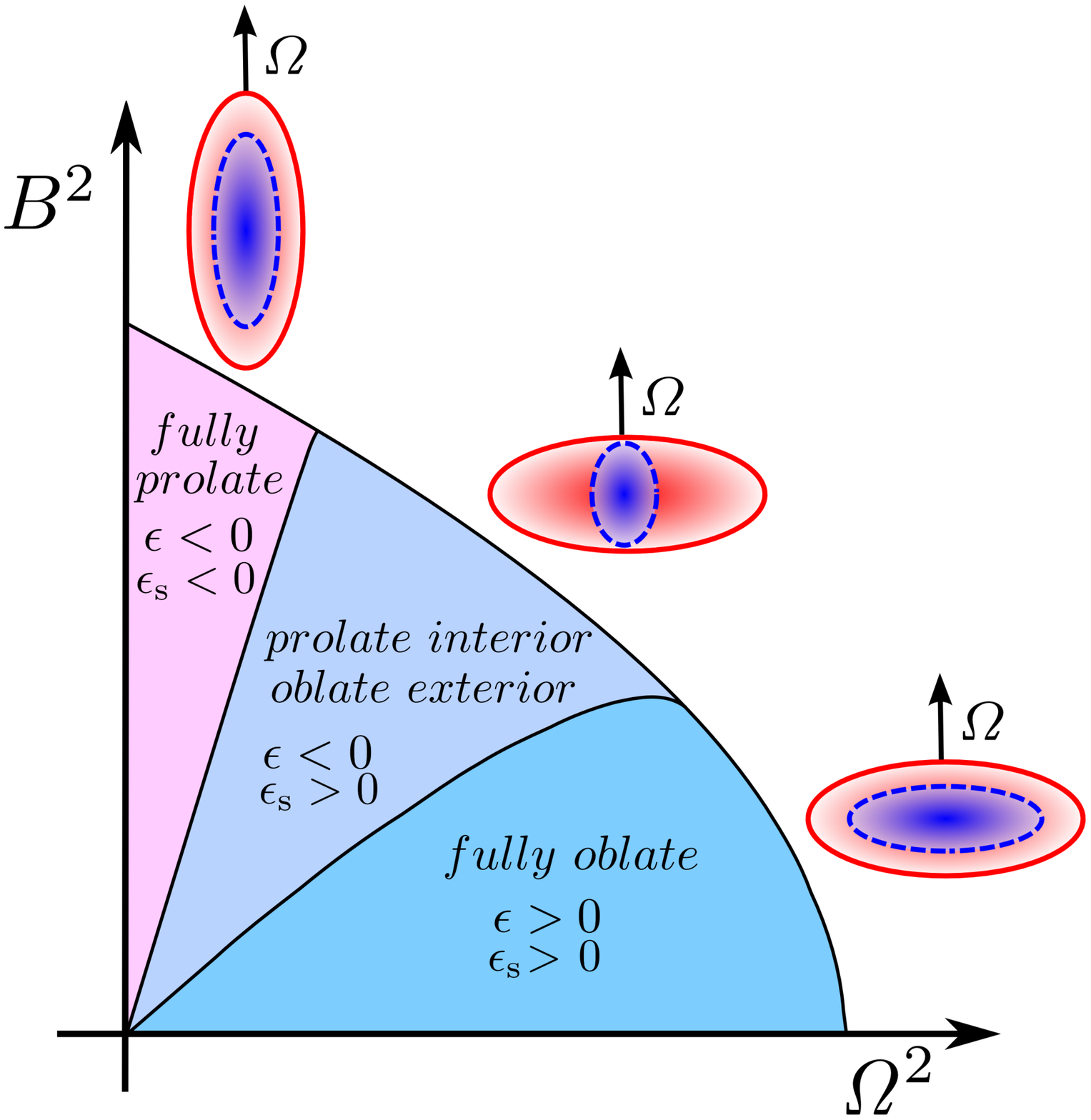,width=2.0in}\figsubcap{a}}
  \hspace*{0.1in}
  \parbox{2.6in}{\epsfig{figure=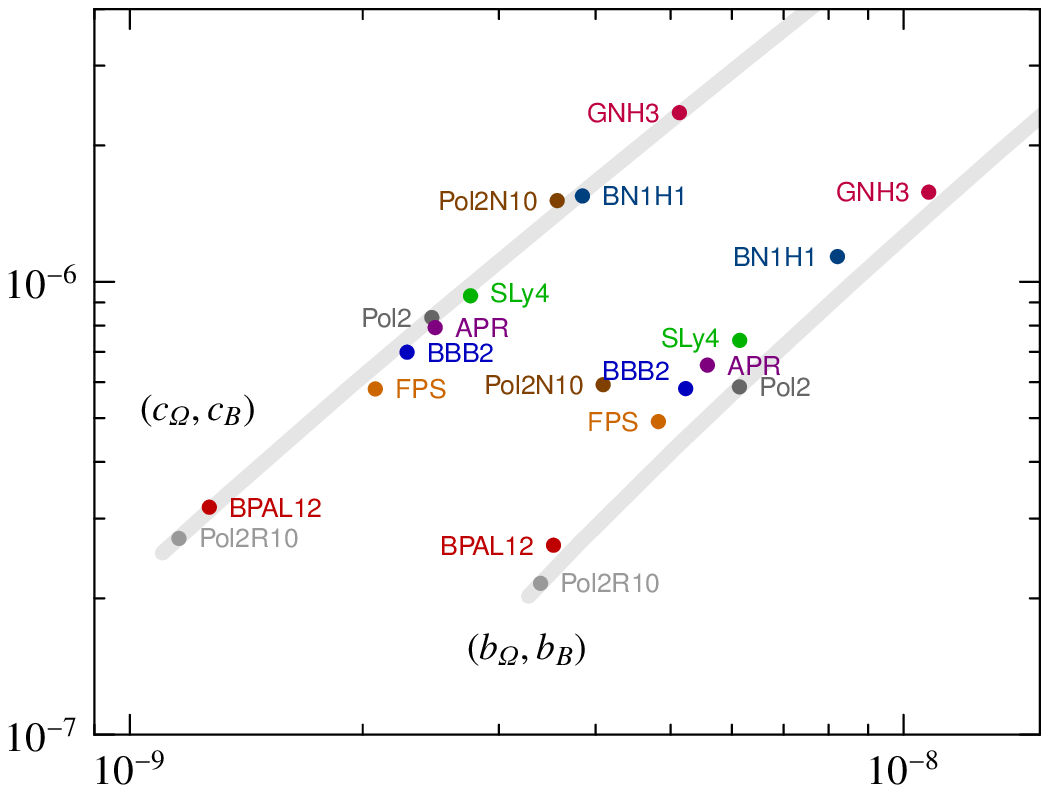,width=2.6in}\figsubcap{b}}
  \caption{(a) Solution space restricted to magnetized and rotating
  Pol2 EOS models between the non-magnetized limit and the
  maximum field strength limit.
  Three distinct classes depending on the relative strength of
  magnetic and centrifugal forces can be distinguished.
  (b) Distortion coefficients $(b_{\mathit{\Omega}},b_{B})$ for the surface
  deformation $\epsilon_{\mathrm{s}}$ and $(c_{\mathit{\Omega}},c_{B})$ for
  the quadrupole distortion $\epsilon$ obtained by perturbing
  non-magnetized and non-rotating models with a gravitational
  mass of $M=1.4 \, M_{\odot}$. In addition, coefficients for a
  Newtonian model {\tt Pol2N10} with $R=10 \, \mathrm{km}$,
  built upon a $\gamma=2$ polytropic EOS, and its relativistic
  counterpart {\tt Pol2R10} are shown. The grey-shaded bands
  correspond to models of increasing circumferential radius $R$
  with a gravitational mass of $M=1.4 \, M_{\odot}$, built with a
  sequence of $\gamma=2$ polytropic EOSs of increasing polytropic
  constant $\kappa$.}
  \label{fig1}
\end{center}
\end{figure}

All models built with a certain EOS have the same rest mass
corresponding to a gravitational mass of $M=1.4 \, M_{\odot}$
in the non-rotating and non-magnetized case. For the polytropic
Pol2 EOS, defined by $p=\kappa \rho^{\gamma}$ with the polytropic
exponent $ \gamma=2$ and the rest-mass density $\rho$, the
adopted polytropic constant $\kappa=83$ (in units in which
$c=G=M_{\odot}=1$) implies a circumferential  radius of
$R=12 \, \mathrm{km}$.

Non-rotating models have been obtained up to large values of
$\lambda_{0}$ (limited only by computational resources) for
all EOSs, and the surface deformation
$\epsilon_{\mathrm{s}} = r_{\mathrm{e}}/r_{\mathrm{p}}-1$,
computed from the equatorial coordinate radius $r_{\mathrm{e}}$
and the polar coordinate radius $r_{\mathrm{p}}$, as well as the
quadrupole distortion $\epsilon = -(3/2) \mathscr{I}_{zz}/{I}$,
obtained from Thorne's quadrupole moment $\mathscr{I}_{zz}$
and the moment of inertia $I$, attain considerable negative
values as the magnetization is increased. The dimensions of
the star even appear to grow without bounds. In turn, the
volume-averaged magnetic field strength
$\langle B^{2} \rangle^{1/2}$ always falls off after attaining
a maximum value of several $10^{17} \, \mathrm{G}$.

The solution space of magnetized and rotating models, parametrized
by $\langle B^{2}\rangle$ and $\mathit{\Omega}^{2}$, has been
determined for the Pol2 EOS. Its lower part up to the maximum field
strength limit, beyond which $\langle B^{2}\rangle$ decreases, is
schematically shown in \fref{fig1}~(a). Since the
curves of vanishing surface deformation, $\epsilon_{\mathrm{s}} = 0$,
and of vanishing quadrupole distortion, $\epsilon = 0$, are different,
the models can be divided into three classes for which surface
deformation and quadrupole distortion are (1) both prolate,
(2) oblate and prolate, or (3) both oblate, depending on the
relative strength of magnetic and centrifugal forces.
In the rotating case, the mass-shedding limit of a magnetized
star is reduced with increasing magnetization in agreement with the
condition of geodesic motion at the stellar equator since the
circumferential equatorial radius is enlarged by the toroidal
magnetic field.

Magnetic field strengths and angular velocities of all known magnetars
are small enough that $\epsilon$ can be well approximated by a linear
function of $\langle B^{2}\rangle$ and $\mathit{\Omega}^{2}$,
$\epsilon = - c_{B} \langle B_{15}^{2}\rangle + c_{\mathit{\Omega}}
\mathit{\Omega}_{0}^{2}$, with the distortion coefficients $c_{B}$ and
$c_{\mathit{\Omega}}$ shown in \fref{fig1}~(b), adopting normalized
variables $B_{15} = B / (10^{15} \, \mathrm{G})$ and
$\mathit{\Omega}_{0} = \mathit{\Omega} / \mathrm{s}^{-1}$.
An estimate for the type II superconducting case\cite{Lander2012}
is then given by $\epsilon = - c_{B} \langle B^{2}_{15}
\rangle^{1/2} \langle B^{2}_{\mathrm{c2},15}
\rangle^{1/2}  + c_{\mathit{\Omega}} \mathit{\Omega}_{0}^{2}$
below the second critical magnetic field strength $\langle B^{2}_{\mathrm{c2}}
\rangle^{1/2} \simeq 7.6 \times 10^{15} \, \mathrm{G} $.
Likewise, $\epsilon_{\mathrm{s}}$ can be computed by using $b_{B}$
and $b_{\mathit{\Omega}}$ instead of $c_{B}$ and $c_{\mathit{\Omega}}$.
The Newtonian model {\tt Pol2N10} with $R=10 \, \mathrm{km}$ and its
relativistic counterpart {\tt Pol2R10} demonstrate that relativistic
effects strongly attenuate both the surface deformation induced by
the toroidal magnetic field and the quadrupole deformation in general.
In contrast, the rotational surface deformation is only slightly
reduced since the centrifugal force is more effective at larger
distances from the rotation axis where relativistic effects have
already weakened.

\section*{Acknowledgments}

This work was supported
in part by the DFG grant SFB/Transregio 7. JF gratefully acknowledges
financial support from the Daimler und Benz Stiftung.

\bibliographystyle{ws-procs975x65}
\bibliography{references}

\end{document}